# Giant optical nonlinearity cancellation in quantum wells

S. Houver[1,5], A. Lebreton[1], T. A. S. Pereira[2], G. Xu[3,6], R. Colombelli[3], I. Kundu[4], L. H. Li[4], E. H. Linfield[4], A. G. Davies[4], J. Mangeney[1], J. Tignon[1], R. Ferreira[1], S. S. Dhillon[1]

[1] Laboratoire Pierre Aigrain, Département de physique de l'ENS, École normale supérieure, PSL Research University, Université Paris Diderot, Sorbonne Paris Cité, Sorbonne Universités, UPMC Univ. Paris 06, CNRS, 75005, Paris, France

[2] Instituto de Física, Universidade Federal de Mato Grosso, 78060-900, Cuiabá Mato Grosso, Brazil

[3] Centre de Nanosciences et de Nanotechnologies, CNRS UMR 9001, Univ. Paris-Sud, Université Paris-Saclay, C2N-Orsay, 91405 Orsay, Cedex, France

[4] School of Electronic and Electrical Engineering, University of Leeds, Leeds, LS2 9JT, UK

[5] Present address: Institute for Quantum Electronics, ETH Zürich, 8093 Zurich, Switzerland

[6] Present address: Key Laboratory of Infrared Imaging Materials and Detectors, Shanghai Institute of Technical Physics, Chinese Academy of Sciences, Shanghai 200083, China

Email: shouver@phys.ethz.ch, sukhdeep.dhillon@lpa.ens.fr

Second-order optical nonlinearities can be greatly enhanced by orders of magnitude in resonantly excited nanostructures, theoretically predicted and experimentally investigated in a variety of semiconductor systems. These resonant nonlinearities continually attract attention, particularly in newly discovered materials, but tend not to be as efficient as currently predicted. This limits their exploitation in frequency conversion. Here, we present a clear-cut theoretical and experimental demonstration that the second-order nonlinear susceptibility can vary by orders of magnitude as a result of giant cancellation effects in systems with many confined quantum states. Using terahertz quantum cascade lasers as a model source to investigate interband and intersubband resonant nonlinearities, we show that these giant cancellations are a result of interfering second-order nonlinear contributions of light and heavy hole states. As well as of importance to understand and engineer the resonant optical properties of materials, this work can be employed as a new, extremely sensitive tool to elucidate the bandstructure properties of complex quantum well systems.

Nonlinear-frequency generation is a ubiquitous technique, which depends critically on the magnitude of the second order nonlinear susceptibility. Strongly enhanced second-order optical nonlinearities, that are orders of magnitude larger than in bulk, have been theoretically investigated since the late 80's [1-4] and experimentally demonstrated in various semiconductor nanostructures under resonant interband or intersubband excitation (for example [5-9]). These resonant nonlinearities are of perpetual interest, especially whenever new materials are discovered such as monolayer transition metal dichalcogenides (TMDs) [10-13]. However, resonant nonlinear interactions involving transitions between semiconductor bands tend to be less efficient than currently predicted, perpetually limiting their uses in nonlinear physics and applications, as well as for frequency conversion.

In this paper, we theoretically and experimentally show the strong interplay of resonant nonlinearities that occur when intersubband and interband transitions are combined through excitation by terahertz (THz) frequency (photons of energy $E_{THz} \sim 10\ meV$) and near infrared (NIR, photons of energy $E_{NIR} \sim 1.5\ eV$) pumps, respectively. The second-order nonlinearity is thus doubly enhanced, permitting efficient *sideband generation* on an optical carrier ($E_{NIR} \pm E_{QCL}$) [5]. Recently it has been shown that quantum cascade lasers (QCLs) can be used as both the source for THz and mid-infrared (MIR) radiation (and intersubband excitation) and the nonlinear medium with an external NIR excitation coupled into the QCL cavity [14-18]. (QCLs exploit intersubband electronic transitions that enable laser action in the THz and MIR regions of the electromagnetic spectrum). Here, we exploit this nonlinear response and demonstrate that the second-order susceptibility, in a complex quantum well nanostructure containing many conduction and valence quantum confined states, presents giant variations with the excitation energy. We highlight for the first time an order of magnitude reduction of the second order nonlinear susceptibility at specific frequencies, particularly important when studying THz (i.e. low energy) transitions. Our theoretical model show that these effects, unpredicted by customarily used and simple three-state models, result from pronounced susceptibility compensations between the numerous nonlinear contributions from the many light (LH) and heavy (HH) hole states. To demonstrate this effect experimentally, a novel reflection geometry has been implemented that permits nonlinear NIR-THz frequency mixing, exciting both LH and HH states. We also show that the nonlinear conversion can be realized over a much larger pump energy range (> 50 meV), when compared to a transmission geometry, and can be used to probe the complex QCL bandstructure.

The nonlinear optical process between a NIR beam and a THz beam leads to the generation of beams at the sum frequency ($E_{sum}=E_{NIR}+E_{THz}$) and difference frequency ($E_{diff}=E_{NIR}-E_{THz}$), as shown schematically in figure 1a. The efficiency of such second-order nonlinear processes depends on the value of the nonlinear susceptibility $\chi^{(2)}$. If the energy of one or both beams is resonant with an existing electronic transition, $\chi^{(2)}$ can be greatly enhanced. The NIR pump in fig. 1a is resonant with an interband transition between a hole and an electron state and the THz beam is resonant with an intersubband transition between two conduction band states.

Previous detailed models of the nonlinear susceptibility [5, 19, 20] do not suit complex structures, such as QCLs which contain a large number of quantum states. We have consequently developed an extended model to calculate the nonlinear susceptibility of a multi-quantum-well structure, by integrating over all possible transitions between electron and hole states. This model is vital to understand the outcome of nonlinear frequency generation when different hole types are excited, as in the case of the excitation geometry schematized in the inset of figure 1a. The NIR excitation propagates along the growth axis and is polarized in the plane of the quantum well layers, which will lead to excitation of both light and heavy hole-electron (LH-el and HH-el) transitions, owing to interband selection rules [21]. The THz polarization is perpendicular to the quantum well layers (i.e. TM polarized). We refer to this excitation configuration as the "reflection geometry" below. It is important to underline that close to resonant excitation, the NIR pump is largely absorbed, rendering the role of phase mismatch negligible [22].

In this study we consider a THz QCL, based on GaAs/Al$_{0.15}$Ga$_{0.85}$As QWs and barrier layers, designed to emit around 3 THz (12 meV, 80 µm, from [23]). Further details on the sample are presented in the Methods. Figure 1b shows the calculated QCL bandstructure diagram with electron states and both light (left) and heavy (right) hole states. Among the large number of electron and hole states per period, those that contribute the most to both the QCL emission and the nonlinear response are highlighted (in colour) that we discuss further below. (The THz emission of the QCL is between electronic (conduction) band states (predominately E5-E4), with the external NIR emission exciting transitions between hole and electronic states).

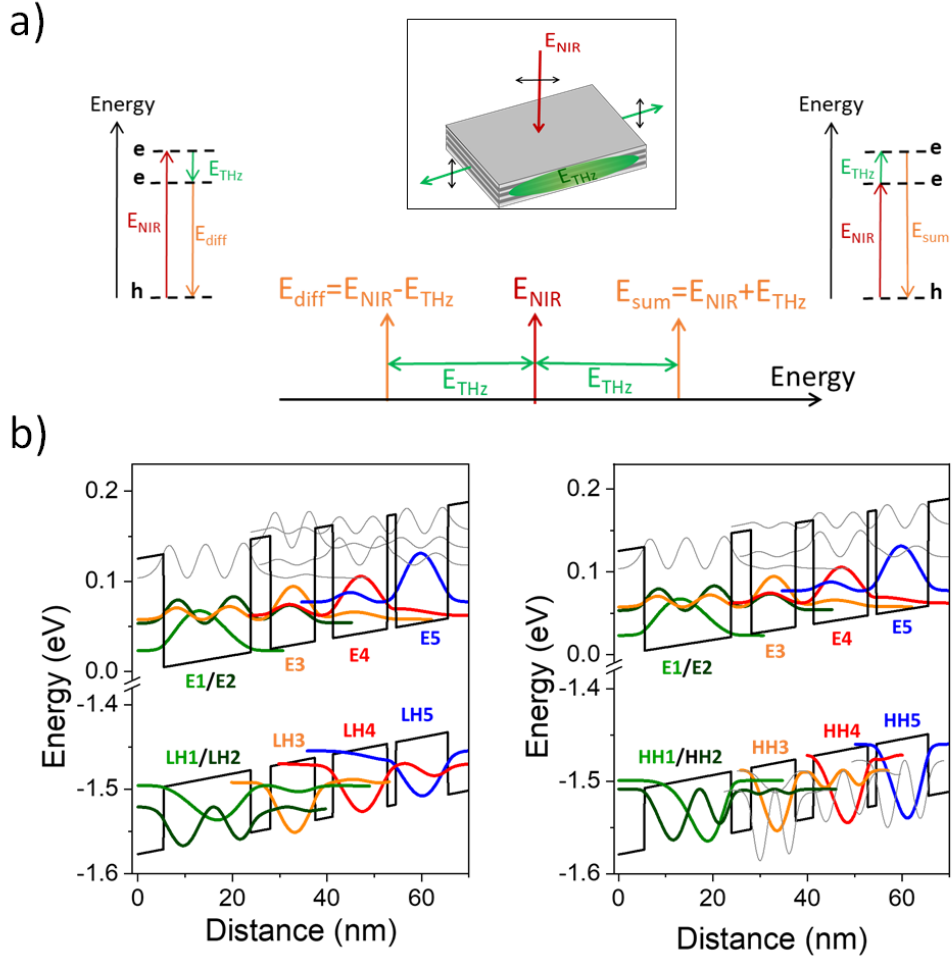

Figure 1: **Resonant nonlinear frequency mixing and quantum cascade laser bandstructure** a) Schematic diagram showing difference (left) and sum (right) frequency generation in a resonant excitation geometry and the corresponding spectrum. The green, red and orange arrows represent the THz, NIR and mixing beams, respectively. Inset: Case employed for resonant excitation, with the external NIR polarization in the plane of the semiconductor QW layers b) Moduli squared of the relevant wavefunctions, shown with corresponding energies for electronic levels in the conduction band potential and for light holes (LH) levels (left) and heavy holes (HH) levels (right) respectively, in the valence band potential. The main electron and hole levels contributing to large interband dipoles are plotted in colours. The electric field applied to the structure for these simulations is 10 kV/cm.

The general nonlinear susceptibility expression for sum frequency generation process $\chi^{(2)}_{sum}$ is given in eq. (1) in the Methods, and how we extend this expression to take into account complex structures. Of particular importance, we point out the sum over all the various confined electron and hole states in the bandstructure. We assume independent light and heavy parabolic dispersions, so that $\chi^{(2)}$ applies separately for either contribution: LH and HH results add algebraically to give the total susceptibility for the reflection geometry. As we demonstrate below, the consideration of both the intra-LH and intra-HH interferences as well those of mixed LH-HH are crucial to explain the full set of results. This is in stark contrast with respect to the linear response, where the absorption intensities add, and cancellation effects do not occur (neither intra nor mixed contributions). As the susceptibility includes all three-state possibilities with one hole and two electron states, and one electron and two hole states, the susceptibility (related to either LH or HH states) can then be split into two terms, $\chi_c^{(2)}$ and $\chi_v^{(2)}$ refering to the conduction and valence bands respectively, as detailed in eq. (2) in the Methods. An essential aspect of this equation is the sign of the triple dipole product in the numerator, which can be positive or negative depending on the transitions involved and can result in giant susceptibility reductions. Implementing the wavefunctions and energies calculated as mentioned previously, the nonlinear susceptibility $\chi_{sum}^{(2)}$ for the QCL structure was determined.

As the nonlinear efficiency is proportional to the square of the nonlinear susceptibility [2, 22], the modulus squared, $|\chi^{(2)}|^2$, is plotted in figure 2a as a function of the sum energy for three configurations: taking into account the HH-el transitions only (in red); the LH-el transitions only (in blue), and both LH-el and HH-el transitions (in orange, corresponding to the reflection geometry). Interestingly, each configuration shows different responses. For energies lower than 1.535 eV, $\left|\chi_{LH}^{(2)}\right|^2$ is significantly lower than $\left|\chi_{HH}^{(2)}\right|^2$. Two features are worth highlighting. First, close to 1.548 eV, both HH and LH contributions display a pronounced minimum, as does the combined signal. Second, close to 1.535 eV none of the HH and LH signals display alone a dip, whilst the combined contribution has a clear minimum. To better understand these effects, the real and imaginary part of $\chi^{(2)}$ (top and bottom parts of figure 2b, respectively) were calculated taking into account the individual contributions of HH-el transitions and LH-el transitions, and their combined values. The energies of the two minima in the $|\chi^{(2)}|^2$ are indicated by green dashed lines. We can thus clearly observe that the minima of the combined signal (orange curve) occur only when both the real and imaginary parts of $\chi^{(2)}$ simultaneously vanish or are close to zero (i.e. only around 1.535 eV and 1.548 eV).

Importantly, we note that for the minimum at 1.535 eV, the real part of the calculated susceptibilities from LH-el only and HH-el only are not close to zero but are of opposite signs: this strongly illustrates the important effect of the susceptibility sign and corresponding cancellation effects, that substantially affect the second-order response of actual multiple QW structures. Similarly, the two imaginary parts in figure 2b further weaken the total susceptibility between the two minima: the orange curve changes sign and remains essentially between the two independent contributions (red and blue curves). This nuanced description comprising cancellations of the nonlinear susceptibility from several positive and negative contributions is missed within a simpler three-state model, as done in earlier works [5, 19, 20].

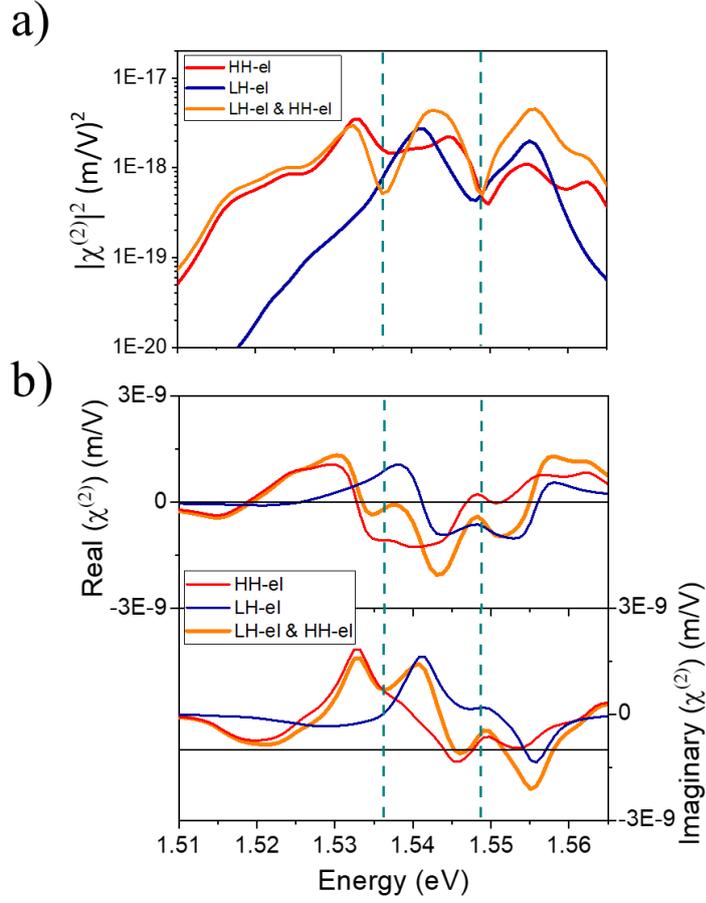

Figure 2: **Minima in the second order nonlinear susceptibility** a) Modulus squared of the second order susceptibility $|\chi^{(2)}|^2$ as a function of sum energy calculated taking into account HH-el transitions only (red curve), LH-el transitions only (blue curve) and calculated for a combination of LH-el and HH-el transitions (orange curve), according to selection rules in reflection geometry. b) Real part (top) and imaginary part (bottom) of calculated $\chi^{(2)}$ for SFG process, considering LH-el only (blue curves); HH-el only (red curves); and, combined (orange curves), as a function of energy. The energy minima in the $|\chi^{(2)}|^2$ combined spectrum, are indicated as green dashed lines and highlight the energies where both real and imaginary total parts (orange) are, or are close to, 0.

To investigate this cancellation of the nonlinear susceptibility experimentally, we implemented a novel excitation geometry within a THz QCL cavity. Previous experimental demonstrations of intracavity frequency mixing in THz QCLs exploited a transmission geometry along the long direction of the ridge. This can only probe the lowest lying interband states as the absorption strongly increases with the interband energy: it simultaneously severely reduces the region where SFG occurs to a very small part of the cavity and suppresses its propagation along the cavity towards the opposite side (absorption length is typically a few microns whereas the cavity is a few millimeters). [14, 15, 17, 18]. To circumvent this limitation and study the nonlinear frequency mixing over a broad range of NIR energies and excite transitions involving both LH and HH states, we adopt a reflection geometry with the NIR excitation incident normal to the QCL ridge (schematized in the inset of figure 1a, similarly used in ref [16] for MIR QCLs). As we will show, this geometry optimizes the ratio between the generation and propagation lengths for the SFG signal, allowing access to a much wider spectral region. As the top of the QCL comprises a metal layer for mode confinement of the TM mode [24], two slits were etched into this metal layer in order to enable NIR propagation into the device and interband excitations in this geometry. The two-slit configuration (1-mm-long 3-µm-wide slits separated by 45 µm) is explained in the Methods.

A schematic diagram of the excitation geometry is shown in figure 3a. The pump beam was focused using a short-focus cylindrical lens creating an elliptical beam shape that covers the area of the slits. The injected NIR beam propagates through the QCL layers, eventually reflected off the bottom gold layer of the QCL and re-crosses the layers before exiting through the metal slit (if not completely absorbed). The sideband signal, which may be generated during either forward or backward paths of the pump, exits the structure through the slit and is finally detected. (The expected paths for the pump and the SFG beams are shown in figure 3b). The experimental set-up is further presented in the Methods. To measure the resonant SFG response, the pump energy $E_{NIR}$ is set below the bandgap then slowly tuned to higher energies. An example of a measured SFG spectrum is shown in figure 3c for a given pump energy: we clearly see that the pump and the SFG traces are separated by the QCL photon energy of about 12 meV (different peaks corresponding to the spectral emission of the QCL).

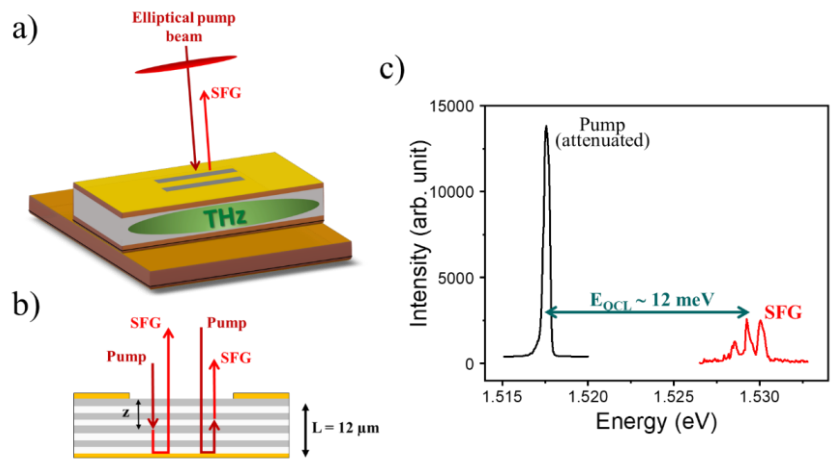

Figure 3: **Sum frequency generation in reflection geometry** a) Schematic diagram of the experimental geometry implemented to excite the QCL. Two slits were etched into the top gold surface of the QCL to let the NIR pump excite the QW structure. b) Propagation of the pump and generated sum beams through the layered structure, the SFG could be generated during either forward or backward path of the pump. c) Example of SFG spectrum around 1.529 eV for a pump excitation near 1.517 eV.

The SFG intensity is plotted on a logarithmic scale as a function of the sum energy (i.e. $E_{sum}=E_{NIR}+E_{THz}$) in figure 4 (black squares, right axis). By significantly reducing the interaction length in the reflection geometry (compared to a transmission geometry), the accessible energy range where the sum frequency is now generated is more than 50 meV, much larger than previous demonstrations of THz QCLs based sideband generation [14, 15]. The intensity increases from lower energies, reaching a maximum around 1.525 eV, then globally decreases until reaching the sensitivity limit at about 1.56 eV. Two significant minima (of about two orders-of-magnitude) occur for sum energies of 1.535 eV and 1.548 eV. The global decrease from the maximum efficiency can be attributed to interband absorption. (Although absorption is significantly reduced in this geometry, it is still important). To account for interband absorption in our model, we consider the factor F detailed in eq. (3) in the Methods. This factor takes into account both the pump and the SFG absorption as they propagate through the structure.

The product $F.|\chi^{(2)}|^2$ is plotted in red in Fig. 4 (left axis), with the same logarithmic scale increment as the experimental data (right axis). Remarkably the SFG intensity and calculated $F.|\chi^{(2)}|^2$ show a very similar behaviour as a function of sum energy, both featuring two significant minima near 1.535 eV and 1.548 eV illustrating the effect of the susceptibility sign. Although the first minimum near 1.535 eV is not as deep in the simulations as in the experimental data and is slightly shifted to 1.538 eV, this discrepancy is most likely owing to small differences between the calculated bandstructure and the actual grown QW structure. Indeed, as discussed below, the calculations are very sensitive to the form of the wavefunction, the number of quantum states considered, the applied field and the effect of band bending due to charge migration, which only change the bandstructure slightly but can considerably affect the nonlinear susceptibility. Note that the minima have not been observed for this QCL in transmission geometries (not shown here), due to the much stronger absorption (also in agreement with modelling, which predicts a much larger F factor).

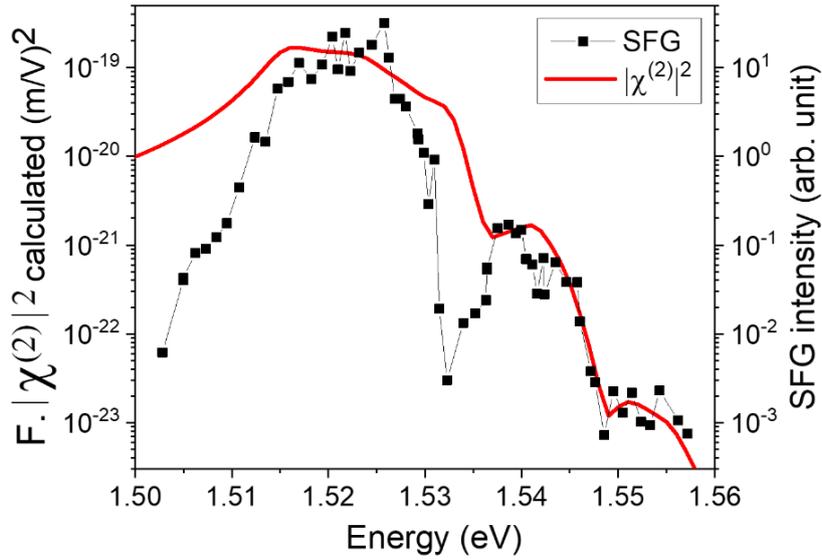

Figure 4: **SFG intensity from experimental data and theoretical model**. SFG intensity (black squares, right axis) measured in the reflection geometry with NIR excitation through the top slits, as a function of sum energy. Modulus squared F.|χ$^{(2)}$|$^2$ (red line, left axis) calculated for the same geometry, taking into account the appropriate combination of LH-el and HH-el, and the absorption factor. Both plots are presented with the same logarithmic scale increment.

To investigate the origin and the importance of these minima further, we have analysed in more detail the interband dipole elements of the numerous LH-el and HH-el transitions. Figure 5a shows the transitions with the largest calculated dipoles. As expected, the largest dipoles (in absolute value) appear for transitions involving electron and hole states with the strongest spatial overlaps (see figure 1b). Note finally the different signs of the dipoles. The different relative signs and intensities in figure 5a assist in understanding the existence of important cancellation effects. It also worth noticing in figure 5a, a concentration of large dipoles between the two minima, as well as the absence of large interband dipoles for energies around the minima (indicated as green dashed lines). As most of the large dipoles result from transitions between similarly confined electron and hole states (En, LHn, HHn with n = 1, 3, 4 and 5 in figure 1b), we can then question whether a reduced model that only considers these dominant interband transitions still permits a good description of the experimental results. In figure 5b, we show calculated |χ$^{(2)}$|$^2$ obtained by taking into account either all possible states (orange curve) or only En, LHn and HHn states ("1st order states", purple curve). We observe that the reduced model still describes the second minimum near 1.548 meV accurately, with an even deeper drop, likely owing to the absence of the nearby large dipoles LH2-E2 and HH1-E2 in this model. However, the first minimum is less visible and blue shifted in the reduced scheme, highlighting the importance of the summing over all states, and its strong and unavoidable cancellation effects on the second order susceptibility in forming an accurate description of the nonlinear interaction.

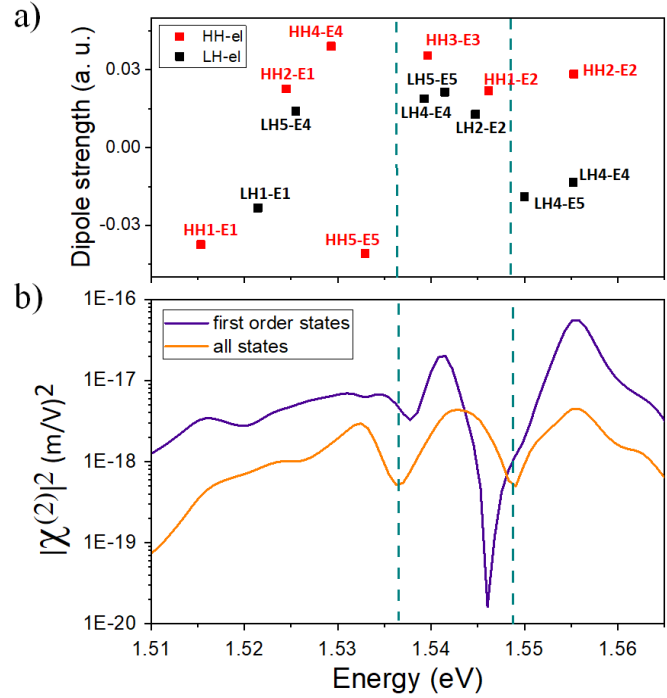

Figure 5: **Susceptibility minima origins** a) Interband dipole strength of both HH-el (red) and LH-el (black) transitions (only dipoles with absolute strength > 0.01 a.u. are shown). b) Modulus squared $|\chi^{(2)}|^2$ calculated in reflection geometry for both HH-el and LH-el transitions, taking into account all states (orange curve) and only the first order confined states (purple curve). Minima previously discussed are indicated by green dashed lines.

It is also worth pointing out that small changes in the parameters of the QCL structure may lead to discernible changes in the nonlinear susceptibility, not only providing important information on the bandstructure but also introducing a mechanism to tune the cancellations in the nonlinear susceptibility. To illustrate this point, the sensitivity of $\chi^{(2)}$ simulations to a small variation of the external electric field applied to the QCL structure (assuming unrealistically that the laser action from the QCL is always present) is shown in figure S3. By changing the field from 7kV/cm to 12 kV/cm, the energy separation of the two dips continuously increases while their relative intensities reverses with increasing field. These two sizeable evolutions demonstrate that both the minima positions and their visibilities are very sensitive to the external applied electric field and hence to the bandstructure profile. This illustrates a scheme to engineer the nonlinear susceptibility, whereby a strong enhancement or suppression can be achieved for certain frequencies by a slight modification of the wavefunctions and applied potential drops. On the other hand, by comparing the SFG intensity experimental data with the complete model could potentially allow to extract the precise bandstructure of complex QW structures such as QCLs, which are otherwise inaccessible by linear optical spectroscopy (as e.g. by absorption, which is insensitive to any cancellation effect, and usually shows an unstructured form in actual QCLs owing to closely spaced energy levels).

To conclude, we have experimentally and theoretically investigated the important interplay of resonant nonlinearities $\chi^{(2)}$ for interband and intersubband excitations in a complex multiple-quantum-well structure. This has highlighted strong deviations from the expected response and shown giant cancellations of the nonlinear susceptibility at certain energies, as well an overall reduction, owing to different contributions from light and heavy holes. This unusual nonlinear susceptibility behaviour was experimentally demonstrated by nonlinear frequency mixing between a NIR and THz beam within a QCL, demonstrating the importance of considering all the quantum states in such a system. This model and experimental observations open the possibility for the careful design and engineering of quantum well bandstructures in order to obtain a desired spectrum of nonlinear susceptibility or to use nonlinear frequency mixing as a probe of complex band structures that are otherwise inaccessible. One can consider engineering structures to achieve a nonlinear response with tuneable minima, potential extinctions or even enhancements at specific energies, adding destructively or constructively contributions from competing contributions. This will further the manipulation of quantum confined states for efficient frequency conversion in nanostructured materials.

## Acknowledgements

We acknowledge financial support from the French National Research Agency (ANR-13-BS03-0001 "RE-LINQ") and from the EPSRC (UK) program grant HyperTerahertz EP/P021859/1. EHL acknowledges support from the Royal Society and the Wolfson Foundation

## Author contributions

S.H. conceived and set up the experiment, acquired and interpreted the experimental data and developed the simulations. A.L. acquired and interpreted the experimental data. S.S.D. conceived the experimental concept. T.A.S.P. and R.F. developed the theoretical model. Sample growth was performed by L.L, E.H.L and A.G.D., sample process was performed by G.X. and R.C. and the slit etching was performed by I.K. The manuscript was written and the data interpreted by S.H., R.F. and S.S.D. R.C, J.M. and J.T. provided insights. All work was coordinated and overseen by S.S.D. All authors contributed to the discussion and to the final manuscript.

# Methods

**QCL bandstructure**

For this study we considered a THz QCL, fabricated from the GaAs/Al$_{0.15}$Ga$_{0.85}$As heterostructure material system. The active region is based on a four QW design relying on a diagonal transition coupled to a phonon extraction stage [23]. It is designed to emit around 3 THz (12 meV, 80 µm). The layer sequence of one period of structure in nanometers is **5.5**/11.0/**1.8**/11.5/**3.8**/9.4/**4.2**/<u>18.4</u> nm. Al$_{0.15}$Ga$_{0.85}$As barriers are indicated in bold font and GaAs wells are in normal font. A Si n-doped GaAs layer (n = 2 × 10$^{16}$ cm$^{-3}$) is underlined. The calculated QCL bandstructure diagram using a Schrödinger solver for electrons and holes states was presented in figure 1b. The laser transition occurs between the upper electronic level E5 (blue line) and lower electronic level E4 (red line). The first bandstructure (on the left) shows electron and light holes (LH) confined states while the second bandstructure shows electron and heavy holes (HH) confined states. The electron-hole dipoles are largest for transitions between first-order confined hole and corresponding electron states, plotted in the same colour in figure 1b (H1-E1 in green, H3-E3 in orange, H4-E4 in red and H5-E5 in blue).

**Second order nonlinear susceptibility model**

The general nonlinear susceptibility expression for sum frequency generation process $\chi^{(2)}_{sum}$ is given by [25]:

$$\chi^{(2)}_{sum} = \frac{1}{\varepsilon_0 V} \int_k \sum_{mnv} \frac{\mu_{mn}\mu_{nv}\mu_{vm}}{\Delta E_{mn} + E_k - E_{sum} - i\Gamma} \left( \frac{\rho_m - \rho_v}{\Delta E_{vm} + E_k - E_{NIR} - i\Gamma} + \frac{\rho_n - \rho_v}{\Delta E_{nv} + E_k - E_{NIR} - i\Gamma} \right) \quad (1)$$

$m,n,v$ are the various confined electron and hole states in the bandstructure and $k$ is the in-plane wavevector (which is conserved in the dipole approximation). $\Delta E_{ij}$ ($\mu_{ij}$) refers to the transition energy (dipole matrix element) between states $i$ and $j$. $E_k$ is the kinetic energy for the relative electron-hole motion, $\rho_i$ represents the population of state $i$. The broadening coefficient $\Gamma$ was set to 2 meV corresponding to a QCL operating temperature of 10 K [26]. $E_{NIR}$ and $E_{sum}$ refer to the NIR pump energy and the sum energy, respectively. The occupancy $\rho_i$ is taken to be 1 for states in the valence band and, owing to the low level doping and weak photo-excitation, is negligible for states in the conduction band. The susceptibility (related to either LH or HH states) can then be split into two terms, $\chi^{(2)}_c$ and $\chi^{(2)}_v$ refering to the conduction and valence bands respectively:

$$\chi^{(2)}_{sum} = \chi^{(2)}_c + \chi^{(2)}_v$$

$$= \frac{1}{\varepsilon_0 V} \int_k \left( \sum_{mnn"} \frac{\mu_{mn}\mu_{nn"}\mu_{n"m}}{(\Delta E_{mn} + E_k - E_{sum} - i\Gamma)(\Delta E_{n"m} + E_k - E_{NIR} - i\Gamma)} \right.$$

$$\left. + \sum_{mm"n} \frac{-\mu_{mn}\mu_{nm"}\mu_{m"m}}{(\Delta E_{nm} + E_k - E_{sum} - i\Gamma)(\Delta E_{nm"} + E_k - E_{NIR} - i\Gamma)} \right) \quad (2)$$

Indices $n$ and $n"$ refer to electron states in the conduction band and indices $m$ and $m"$ refer to hole states in the valence band.

Note that $|\chi^{(2)}|^2$, calculated at 10 K, has been red shifted by 13 meV to account for a bandgap shift, observed in the experiments, due to a local temperature increase by the electrical power dissipated in the QCL [27]. This shift appears clearly when comparing (see figure S2) the photoluminescence of the biased QCL with the calculated energies (at 10 K) of the first electron-hole transitions of the structure.

**QCL fabrication and slit design**

The THz QCL structure was grown by molecular beam epitaxy. The active region comprises 180 periods resulting in a total thickness of 12 μm. The THz QCL was processed into a metal-metal waveguide, where the active region is sandwiched between two metal stripes, with 150-μm-wide ridges, cleaved into 3-mm-long cavities and indium soldered to copper mounts. As the top of the QCL comprises a metal layer for mode confinement, apertures need to be etched into this metal layer for the NIR to propagate into the device and enable interband excitations in this geometry. The aperture positions were judiciously designed via numerical simulations of the THz field intensity in the QCL cavity such that the QCL performance is not affected, and there is an overlap between the interacting NIR and THz electric fields. Results of these simulations are presented in figure S1 in the Sup. Mat., where a two-slit configuration (1-mm-long 3-μm-wide slits separated by 45 μm) is chosen since the fundamental and first excited modes are found to have a significant field under the apertures (see figure 3a). The two-slit configuration was realized by focused ion-beam etching. The QCL output power and voltage-current characteristics were found to be very similar after etching the slits to those before etching. We note that the emitted angle of the generated NIR beam is expected to be similar to that of the NIR pump since the THz k-vector is small compared to that of the NIR.

**Experimental setup**

The device was placed in a continuous flow cryostat at 10 K. The external NIR excitation was provided by a continuous wave tuneable titanium-sapphire laser with an output power set to 2 mW. Although all polarizations of the NIR pump are equivalent for the SFG process (polarization in the plane of the semiconductor layers), we chose a NIR polarization parallel to the slit direction to avoid plasmonic effects on the metal edges. The sideband signal, which may be generated during either forward or backward paths of the pump, is detected using a grating spectrometer coupled to a CCD array.

**Absorption factor**

To account for interband absorption in our model, we consider the factor:

$$\begin{aligned} F &= \int_0^{2L} \exp(-\alpha_P z)\exp(-\alpha_{SFG}(2L-z))\frac{dz}{2L} \\ &= \exp(-2\alpha_P L)\frac{1-\exp(-2(\alpha_{SFG}-\alpha_P)L)}{2L(\alpha_{SFG}-\alpha_P)} \end{aligned} \quad (3)$$

where $\alpha_P$ and $\alpha_{SFG}$ are the energy-dependent absorption coefficients of the pump and SGF, respectively, and L = 12 μm is the sample thickness. This factor takes into account the absorption of the pump (first term on the integrand) and the SFG (second term on the integrand) beams, in either forward (z from 0 to L) and backwards (z from L to 2L) direction of the NIR excitation through the QCL as shown schematically in figure 3b.